\documentclass[aps,prd,showpacs,twocolumn]{revtex4}
\usepackage{amssymb}
\usepackage{amsmath}

\begin{document}

\title{Generalized virial theorem in $f(R)$ gravity}

\author{Christian G.~B\"ohmer}
\email{c.boehmer@ucl.ac.uk} 
\affiliation{Department of Mathematics, University College London,
             Gower Street, London, WC1E 6BT, UK}

\author{Tiberiu Harko}
\email{harko@hkucc.hku.hk} 
\affiliation{Department of Physics and Center for Theoretical
             and Computational Physics, The University of Hong Kong,
             Pok Fu Lam Road, Hong Kong}

\author{Francisco S.~N.~Lobo}
\email{francisco.lobo@port.ac.uk} 
\affiliation{Institute of Cosmology \& Gravitation,
             University of Portsmouth, Portsmouth PO1 2EG, UK}
\affiliation{Centro de Astronomia e Astrof\'{\i}sica da
             Universidade de Lisboa, Campo Grande, Ed. C8 1749-016 Lisboa,
             Portugal}

\date{\today}

\begin{abstract}
We generalize the virial theorem in $f(R)$ modified gravity using the collisionless Boltzmann equation. We find supplementary geometric terms in the modified Einstein equation providing an effective contribution to the gravitational energy. The total virial mass is proportional to the effective mass associated with the new geometrical term, which may account for the well-known virial theorem mass discrepancy in clusters of galaxies. The model predicts that the geometric mass and its effects extend beyond the virial radius of the clusters. We also consider the behavior of the galaxy cluster velocity dispersion in $f(R)$ models. The metric inside the galactic cluster, as well as the Lagrangian of the modified gravity model, are obtained in terms of quantities directly related to the physical properties of the clusters, and which can be determined from astrophysical observations. Thus, the $f(R)$ virial theorem can be an efficient tool in observationally testing the viability of this class of generalized gravity models.
\end{abstract}


\maketitle

\section{Introduction}
\label{sec:a}

The issue of dark matter is a long outstanding problem in modern astrophysics. Two observational aspects, namely, the behavior of the galactic rotation curves and the mass discrepancy in clusters of galaxies led to the necessity of considering the existence of dark matter at a galactic and extra-galactic scale. The rotation curves of spiral galaxies show that the rotational velocities increase from the center of the galaxy and then attain an approximately constant value, $v_{tg\infty} \sim 200-300 {\rm km}/{\rm s}$, within a distance $r$ from the center of the galaxy~\cite{Bi87}. In these regions the mass increases linearly with the radius, even where very little luminous matter can be detected. Relatively to the mass discrepancy in clusters of galaxies, the total mass of a cluster can be estimated in two ways. First, by taking into account the motions of its member galaxies, the virial theorem provides an estimate, $M_{V}$. Second, the total baryonic mass $M$ may be estimated by considering the total sum of each individual member's mass. The mass discrepancy arises as one generally verifies that $M_{V}$ is considerably greater than $M$, with typical values of $M_{V}/M \sim 20-30$~\cite{Bi87}.

This is usually explained by postulating the existence of a dark matter, assumed to be a cold pressure-less medium distributed in a spherical halo around the galaxies. A number of candidates for dark matter have been proposed in the literature, in particular, the most popular ones being the weakly interacting massive particles (WIMP)~\cite{OvWe04}. Their interaction cross section with normal baryonic matter, while extremely small, is expected to be non-zero and therefore may be directly detectable. However, despite more than 20 years of intense experimental and observational research, the \textit{non-gravitational} evidence for dark matter is still lacking. Moreover, accelerator and reactor experiments do not yet support the physics beyond the standard model, on which the dark matter hypothesis is based upon. It is interesting to note that dark matter consisting of WIMPs may exist in the form of an Einstein cluster~\cite{Ein}, or could possibly undergo a phase transition to form a Bose-Einstein condensate~\cite{Bo}. It has also been suggested that the dark matter in the Universe might be composed of super-heavy particles, with mass $\geq 10^{10} {\rm GeV}$. But observational results show that dark matter can be composed of super-heavy particles only if these interact weakly with normal matter or if their mass is above $10^{15} {\rm GeV}$~\cite{AlBa03}.

Therefore, it seems that the possibility that Einstein's (and Newton's) gravity breaks down at the scale of galaxies cannot be excluded \textit{a priori}, and several theoretical models, have been proposed in the literature~\cite{dark}. One possibility is to assume that at large scales general relativity breaks down, and a more general action describes the gravitational field. The theoretical models in which the standard Einstein-Hilbert action can be replaced by an arbitrary function of the Ricci scalar $R$ have recently been extensively investigated~\cite{Bu70}. In this context, cosmic acceleration can arise due to corrections to the usual gravitational action of general relativity in the form $R+1/R$, as now the $1/R$ term dominates for low curvatures~\cite{Carroll:2003wy}. Therefore, modified gravity, in principle, eliminates the need for dark energy, and since a modification of the Einstein-Hilbert action of the form $f(R)\propto R+R^2$ can lead to early-time inflation~\cite{Starobinsky:1980te}, it should also be possible to unify the early and the late time accelerating phases of the Universe.

However, it was shown that in all $f(R)$ theories that behave as a power of $R$, at large or small $R$, the scale factor during the matter phase grows as $t^{1/2}$ instead of the standard law $t^{2/3}$~\cite{Amendola1}. This behavior, which is also a general property of models with $d^2f/dR^2<0$, is inconsistent with cosmological observations (e.g.~WMAP), thereby ruling out these models, even if they pass the supernovae test and can escape the local gravity constraints. Nevertheless, $f(R)$ modified theories of gravity can, in general, give rise to cosmological viable models compatible with a matter-dominated epoch evolving into a late accelerated phase~\cite{Capozziello,odin}. Thus, the viability of the $f(R)$ models proposed has been a fundamental issue extensively analyzed in the literature~\cite{Amendola1,Capozziello,viablemodels,Hu:2007nk,energycond,Nojiri:2006ri}. In this context, severe weak field constraints in the solar system range seem to rule out most of the models proposed so far~\cite{solartests,Chiba,Olmo07}, although viable models do exist~\cite{Hu:2007nk,solartests2,Sawicki:2007tf,Amendola:2007nt}.

Specific models of $f(R)$ gravity that admit a modified Schwarzschild-de Sitter metric and describe both the Pioneer anomaly and the flat rotation curves of spiral galaxies were analyzed in~\cite{Sob,SaRa07}. In particular, within the framework of $f(R)$ gravity, a model has been proposed exhibiting an explicit coupling of an arbitrary function of $R$ with the matter Lagrangian density~\cite{Bertolami:2007gv}, which establishes a connection between the problem of the rotation curve of galaxies, via a solution somewhat similar to the one put forward in the context of MOND, and the Pioneer anomaly. The possibility that the galactic dynamics of massive test particles may be understood without the need for dark matter was also considered in the framework of $f(R)$ gravity~\cite{Cap2,Borowiec:2006qr,Mar1}. The results obtained in these papers seem to suggest that a strong modification of standard general relativity is required to explain the observed behavior of the galactic rotation curves. However, it was shown that to explain the motion of test particles around galaxies, in the constant rotational velocity regions, only mild deviations from classical general relativity are required~\cite{Boehmer:2007kx}.

Due to its generality and wide range of applications, the virial theorem plays an important role in astrophysics. Assuming steady state, one of the important results which can be obtained with the use of the virial theorem is to deduce the mean density of astrophysical objects such as galaxies, clusters and super clusters, by observing the velocities of test particles rotating around them. Hence the virial theorem can be used to predict the total mass of the clusters of galaxies. The virial theorem is also a powerful tool for stability studies. In a general relativistic framework several versions of the virial theorem have been proposed~\cite{vir}, including the effect of a cosmological constant~\cite{Ja70,No}, and the generalization to brane world models~\cite{HaCh07}.

It is the purpose of the present paper to consider the virial theorem in the framework of $f(R)$ modified theories of gravity. Using the collisionless Boltzmann equation in the modified Einstein field equations we derive a generalized virial equality. It takes into account the presence of the supplementary geometric effects due to the modification of the gravitational action.

This supplementary geometric terms give an effective contribution to the gravitational energy, the total virial mass being proportional to the effective mass associated to the new geometrical terms. Therefore this term may account for the well-known virial theorem mass discrepancy in clusters of galaxies. By using the virial theorem together with the gravitational field equations of the $f(R)$ model the metric inside the cluster of galaxies can be obtained in a simple form, in terms of physical parameters that can be fully obtained from observations, like, for example, the temperature of the intra-cluster gas and the radius and central density of the cluster core. Once the metric is known, the Lagrangian of the modified $f(R)$ gravity model can also be obtained in terms of observable physical quantities. To explain the observational data at the extra-galactic scale a logarithmic modification of the standard general relativistic action is needed. Therefore the generalized virial theorem in $f(R)$ gravity can be an efficient tool in observationally testing the viability of this class of generalized gravity models.

This paper is outlined in the following manner. In Section~\ref{sec:b}, we present the gravitational field equations of the galactic clusters in the scalar-tensor version of $f(R)$ modified theories of gravity. The relativistic Boltzmann equation is outlined and the respective generalized virial theorem in $f(R)$ gravity is deduced in Section~\ref{sec:c}.  Astrophysical applications are explored in Section~\ref{sec:d}, in particular, predictions of the geometric mass and geometric radius from galactic cluster observations are presented, and the behavior of the galaxy cluster velocity dispersion in $f(R)$ models is also considered. In Section~\ref{lagr} we derive the Lagrangian of the modified gravity model, and analyze some of its properties. Finally, in Section~\ref{sec:e}, we discuss and conclude our results.

\section{Galactic clusters in the scalar-tensor version of $f(R)$ gravity}
\label{sec:b}

Astronomical observations have shown that galaxies tend to concentrate in
larger structures, called clusters of galaxies. The total mass of galaxy
clusters ranges from $10^{13}$ $M_{\odot }$ for groups up to a few $10^{15}$
$M_{\odot }$ for very rich systems, and the cluster morphology is usually
dominated by a regular centrally peaked main component~\cite{ReBo02,Ar05}.
Since clusters are ``dark matter'' dominated objects their formation and
evolution is driven by gravity. In this context, the mass function of the
clusters is determined by the initial conditions of the mass distribution
set in the early universe, and the evolution of the large scale matter
distribution on scales comparable to the size of the clusters is linear. The
overall process of the gravitational growth of the density fluctuations and
the development of gravitational instabilities leading to cluster formation
has been extensively studied by using both analytical and numerical methods~\cite{Sch01}.

In order to derive the generalization of the relativistic virial theorem for
galaxy clusters in the $f(R)$ gravity models we need, as a first step, to
obtain the gravitational field equations for a static and spherically
symmetric system, with the matter energy-momentum tensor corresponding to a
system of identical, collisionless point particles in random motion. To
derive the basic field equations we will use the scalar-tensor
representation of $f(R)$ gravity, which allows a clear physical
interpretation of the model.

\subsection{Scalar-tensor representation for $f(R)$ gravity}

The action for $f(R)$ modified theories of gravity takes the form
\begin{align}
S=\int \left[\frac{1}{16\pi G}f(R)+L_m\right] \sqrt{-g}\; d^{4}x,
\end{align}
where $f(R)$ is an arbitrary analytical function of the Ricci scalar $R$ and
$L_m$ is the Lagrangian density corresponding to matter.

Varying the action with respect to the metric $g_{\mu \nu }$ yields the
following field equations
\begin{align}
FR_{\mu \nu }-\frac{1}{2}fg_{\mu \nu }-\left(\nabla _{\mu }\nabla _{\nu
}-g_{\mu \nu }\square \right)F=8\pi G \,T_{\mu \nu},  \label{field}
\end{align}
where we have denoted $F=df/d R$ and $T_{\mu \nu}$ is the standard minimally
coupled matter energy-momentum tensor. Note that the covariant derivative of
the field equations and of the matter energy-momentum tensor vanishes for
all $f(R)$ by means of generalized Bianchi identities~\cite{Bertolami:2007gv,Koivisto}. Contracting the field equations gives the useful relation
\begin{align}  \label{trace}
3\square F+FR-2f=8\pi G\,T,
\end{align}
from which one verifies that the Ricci scalar is now a fully dynamical
degree of freedom, and where $T$ is the trace of the energy-momentum tensor.

By introducing the Legendre transformation $\left\{ R,f\right\} \rightarrow
\left\{ \phi ,V\right\} $ defined as
\begin{align}
\phi \equiv F\left( R\right), \quad V\left( \phi \right) \equiv R\left( \phi
\right) F-f\left( R\left( \phi \right) \right) ,
\end{align}
the field equations of $f(R)$ gravity can be reformulated as~\cite{equiv}
\begin{align}
R_{\mu \nu }-\frac{1}{2}g_{\mu \nu }R=8\pi \frac{G}{\phi }T_{\mu \nu
}+\theta _{\mu \nu },  \label{fieldin}
\end{align}
where
\begin{align}
\theta _{\mu \nu }=-\frac{1}{2}V\left( \phi \right) g_{\mu \nu }+\frac{1}{%
\phi }\left( \nabla _{\mu }\nabla _{\nu }-g_{\mu \nu }\square \right) \phi.
\end{align}

Using these variables, Eq.~(\ref{trace}) takes the form
\begin{align}  \label{trace1}
3 \square \phi + 2V(\phi) -\phi \frac{dV}{d\phi }=8\pi GT.
\end{align}

In this representation the field equations of $f(R)$ gravity can be derived
from a Brans-Dicke type action with parameter $\omega =0$, given by
\begin{align}
S=\frac{1}{16\pi G}\int \left[ \phi R-V(\phi) +L_{m}\right] \sqrt{-g}\;
d^{4}x.
\end{align}

The only requirement for the $f(R)$ model equations to be expressed in the
form of a Brans-Dicke theory is that $F(R)$ be invertible, that is, $R(F)$
exists~\cite{Olmo07}. This condition is necessary for the construction of $%
V(\phi)$.

The modification of the standard Einstein-Hilbert action leads to the
appearance in the field equations of an effective gravitational constant $G_{%
\mathrm{eff}}=G/\phi$, which is a function of the curvature. Secondly, a new
source term for the gravitational field, given by the tensor $\theta_{\mu
\nu}$, is also induced. The tensor $\theta_{\mu \nu}$ is determined by the
trace of the energy-momentum tensor via Eq.~(\ref{trace1}), which thus acts
as an independent physical parameter determining the metric of the
space-time.

\subsection{Field equations for a system of identical and collisionless
point particles}

Let us assume a time-oriented Lorentzian four-dimensional space-time
manifold $\mathcal{M}$. Consider an isolated and spherically symmetric
cluster described by a static and spherically symmetric metric
\begin{align}
ds^{2}=-e^{\nu \left( r\right) }dt^{2}+e^{\lambda \left( r\right)
}dr^{2}+r^{2}\left( d\theta ^{2}+\sin^{2}\negmedspace\theta d\varphi
^{2}\right).  \label{line}
\end{align}

The galaxies, which are treated as identical and collisionless point
particles, are described by a distribution function $f_B$. The latter
distribution function obeys the general relativistic Boltzmann equation,
which shall be treated in detail below.

In terms of the distribution function the energy-momentum tensor of the
matter can be written as~\cite{Li66}
\begin{align}
T_{\mu \nu }=\int f_B\, m\, u_{\mu }u_{\nu }\;du,
\end{align}
where $m$ is the mass of the particle (galaxy), $u_{\mu }$ is the
four-velocity of the galaxy and $du =du_{r}du_{\theta }du_{\varphi }/u_{t}$
is the invariant volume element of the velocity space. Therefore, the
energy-momentum tensor of the matter in a cluster of galaxies can be
represented in terms of an effective density $\rho_{\mathrm{eff}}$ and of an
effective anisotropic pressure, with radial $p_{\mathrm{eff}}^{(r)}$ and
tangential $p_{\mathrm{eff}}^{(\perp)}$ components, given by
\begin{align}
\rho_{\mathrm{eff}} &= \rho \left\langle u_{t}^{2}\right\rangle,\quad p_{%
\mathrm{eff}}^{(r)}=\rho \left\langle u_{r}^{2}\right\rangle,  \notag \\
p_{\mathrm{eff}}^{(\perp)} &= \rho \left\langle u_{\theta
}^{2}\right\rangle= \rho \left\langle u_{\varphi }^{2}\right\rangle,
\end{align}
where, at each point, $\left\langle u_{r}^{2}\right\rangle $ is the average
value of $u_{r}^{2}$, etc, and $\rho $ is the mass density~\cite{Ja70}.

By using this form of the energy-momentum tensor, the gravitational field
equations describing a cluster of galaxies, in $f(R)$ gravity, take the form
\begin{multline}
-e^{-\lambda }\left( \frac{1}{r^{2}}-\frac{\lambda ^{\prime }}{r}\right) +
\frac{1}{r^{2}}=8\pi \frac{G}{\phi }\rho \left\langle u_{t}^{2}\right\rangle
\\
-\frac{1}{2\phi }V\left( \phi \right) +\frac{1}{\phi }\left( \nabla
_{t}\nabla ^{t}-\square \right) \phi ,  \label{f1}
\end{multline}
\begin{multline}
e^{-\lambda }\left( \frac{\nu ^{\prime }}{r}+\frac{1}{r^{2}}\right) -\frac{1
}{r^{2}}=8\pi \frac{G}{\phi }\rho \left\langle u_{r}^{2}\right\rangle \\
+\frac{1}{2\phi }V\left( \phi \right) -\frac{1}{\phi }\left( \nabla
_{r}\nabla ^{r}-\square \right) \phi ,  \label{f2}
\end{multline}
\begin{align}
\frac{1}{2}&e^{-\lambda }\left( \nu ^{\prime \prime }+\frac{\nu ^{\prime 2}}{%
2}+\frac{\nu ^{\prime }-\lambda ^{\prime }}{r}-\frac{\nu ^{\prime }\lambda
^{\prime }}{2}\right)  \notag \\
&=8\pi \frac{G}{\phi }\rho \left\langle u_{\theta }^{2}\right\rangle +\frac{1%
}{2\phi }V\left( \phi \right) -\frac{1}{\phi }\left( \nabla _{\theta }\nabla
^{\theta }-\square \right) \phi \\
&=8\pi \frac{G}{\phi }\rho \left\langle u_{\varphi }^{2}\right\rangle +\frac{%
1}{2\phi }V\left( \phi \right) -\frac{1}{\phi }\left( \nabla _{\varphi
}\nabla ^{\varphi }-\square \right) \phi.  \label{f3}
\end{align}

A useful relationship is obtained by adding the gravitational field
equations Eqs.~(\ref{f1})--(\ref{f3}), from which we obtain the following
equation
\begin{multline}
e^{-\lambda }\left( \frac{\nu ^{\prime \prime }}{2}+\frac{\nu ^{\prime 2}}{4}
+\frac{\nu ^{\prime }}{r}-\frac{\nu ^{\prime }\lambda ^{\prime }}{4}\right)
=4\pi \frac{G}{\phi }\rho \left\langle u^{2}\right\rangle \\
+\frac{1}{\phi } V\left( \phi \right) +\frac{1}{\phi }\left( 2\nabla
_{t}\nabla ^{t}+\square \right) \phi ,  \label{ff}
\end{multline}
where $\langle u^{2}\rangle =\langle u_{t}^{2}\rangle +\langle
u_{r}^{2}\rangle +\langle u_{\theta }^{2}\rangle +\langle u_{\varphi
}^{2}\rangle$.

Since we are interested in astrophysical applications at the extra-galactic
level, we may assume that the deviations from standard general relativity
(corresponding to the background value $\phi =1$) are small. Therefore we
may represent $\phi $ as $\phi =1+\epsilon g^{\prime }(R)$, where $\epsilon $
is a small quantity, and $g^{\prime }(R)$ describes the modifications of the
geometry due to the presence of the tensor $\theta _{\mu \nu }$~\cite{Olmo07}%
. Consequently $1/\phi \simeq 1-\epsilon g^{\prime }(R)$, and Eq.~(\ref{ff})
can be rewritten as
\begin{align}
e^{-\lambda }\left( \frac{\nu ^{\prime \prime }}{2}+\frac{\nu ^{\prime 2}}{4}%
+\frac{\nu ^{\prime }}{r}-\frac{\nu ^{\prime }\lambda ^{\prime }}{4}\right)
\simeq 4\pi G\rho \left\langle u^{2}\right\rangle +4\pi G\rho _{\phi },
\label{ff1}
\end{align}
where
\begin{multline}\label{approx}
4\pi G\rho _{\phi } \simeq -4\pi G\epsilon \rho \left\langle
u^{2}\right\rangle g^{\prime }(R) \\
+\left[\frac{1}{\phi }V\left( \phi \right) +\left.\frac{1}{\phi
}\left( 2\nabla _{t}\nabla ^{t}+\square \right) \phi
\right]\right| _{\phi =1+\epsilon g^{\prime }(R)}.
\end{multline}

\section{The virial theorem in $f(R)$ generalized gravity models}

\label{sec:c}

In order to derive the virial theorem for galaxy clusters, which are
described by the distribution function $h$, we have to first write down the
Boltzmann equation governing the evolution of the distribution function.
This equation can then be integrated over the velocity space, to yield an
equation which, used in conjunction with the gravitational field equations,
provides the required generalization of the virial theorem.

\subsection{The relativistic Boltzmann equation}

Consider a tangent bundle $T(\mathcal{M})$, which is a real vector bundle
whose fibers at a point $x\in \mathcal{M}$ is given by the tangent space $%
T_{x}\left(\mathcal{M}\right)$. In the space-time $\mathcal{M}$ the
instantaneous state of a particle with mass $m_0$ is given by a
four-momentum $p\in T_{x}\left( \mathcal{M}\right) $ at an event $x\in
\mathcal{M}$. The one-particle phase space $P_{\mathrm{phase}}$ is a subset
of the tangent bundle given by~\cite{Li66}
\begin{align}
P_{\mathrm{phase}}:=\left\{ \left( x,p\right)\left|\right. x\in {\mathcal{M}}%
,p\in T_{x}\left({\mathcal{M}}\right) ,p^{2}=-m_0^2\right\}.
\end{align}

A state of a multi-particle system is described by a continuous,
non-negative distribution function $f_B(x,p) $, defined on $P_{\mathrm{phase}%
}$, and which gives the number $dN$ of the particles of the system which
cross a certain space-like volume $dV$ at $x$, and whose 4-momenta $p$ lie
within a corresponding three-surface element $d\vec{p}$ in the momentum
space. The mean value of $f_B$ gives the average number of occupied particle
states $(x,p) $. Macroscopic, observable quantities can be defined as
moments of $f_B$~\cite{Li66}.

Let $\{ x^{\alpha }\} $, $\alpha =0,1,2,3$ be local coordinates in an open
set $U\subset\mathcal{M}$. The coordinates are chosen so that $\partial _t$
is time-like future directed and $\partial _a$, $a=1,2,3$, are spacelike.
Then $\{\partial /\partial x^{\alpha }\}$ is the corresponding natural basis
for tangent vectors. We express each tangent vector $p$ in $U$ in terms of
this basis as $p=p^{\alpha }\partial /\partial x^{\alpha }$ and define a
system of local coordinates $\left\{ z^{A}\right\} $, $A=0,\ldots,7$ in 
$T_{U}(\mathcal{M})$ as $z^{\alpha }=x^{\alpha }$, $z^{\alpha +4}=p^{\alpha }$.
This defines a natural basis in the tangent space given by $\left\{ \partial
/\partial z^{A}\right\} =\left\{ \partial /\partial x^{\alpha },\partial
/\partial p^{\alpha }\right\} $~\cite{Li66}.

A vertical vector field over $T\left({\mathcal{M}}\right)$ is given by $\pi
=p^{\alpha }\partial /\partial p^{\alpha }$. The geodesic flow field $\sigma
$, which can be constructed over the tangent bundle, is defined as $\sigma
=p^{\alpha }\partial /\partial x^{\alpha }-p^{\alpha }p^{\gamma }\Gamma
_{\alpha \gamma }^{\beta }\partial /\partial p^{\beta }=p^{\alpha }D_{\alpha
}$, where $\Gamma _{\alpha \gamma }^{\beta }$ are the connection
coefficients. Physically, $\sigma $ describes the phase flow for a stream of
particles whose motion through space-time is geodesic~\cite{Li66}.

Therefore, the transport equation for the propagation of a particle in a
curved arbitrary Riemannian space-time is given by the Boltzmann equation~\cite{Li66}
\begin{align}  \label{distr}
\left( p^{\alpha }\frac{\partial }{\partial x^{\alpha }}-p^{\alpha }p^{\beta
}\Gamma _{\alpha \beta }^{i}\frac{\partial }{\partial p^{i}}\right) f_B=0.
\end{align}

For many applications it is convenient to introduce an appropriate
orthonormal frame or tetrad $e_{\mu }^{a}(x)$, $a=0,1,2,3$, which varies
smoothly over some coordinates neighborhood $U$ and satisfies the condition $%
g^{\mu\nu} e_{\mu}^{a} e_{\nu}^{b} =\eta^{ab}$ for all $x\in U$~\cite
{Li66,Ja70}. Any tangent vector $p^{\mu}$ at $x$ can be expressed as $%
p^{\mu}=p^{a}e_{a}^{\mu}$, which defines the tetrad components $p^{a}$.

In the case of the spherically symmetric line element given by Eq.~(\ref
{line}) we introduce the following frame of orthonormal vectors~\cite
{Li66,Ja70}:
\begin{alignat}{2}
e_{\mu}^{0} &= e^{\nu /2}\delta _{\mu}^{0}, & \qquad e_{\mu}^{1} &=
e^{\lambda /2}\delta _{\mu }^{1},  \notag \\
e_{\mu}^{2} &= r\delta_{\mu}^{2}, & \qquad e_{\mu}^{3} &= r\sin \theta
\delta_{\mu}^{3}.
\end{alignat}

Let $u^{\mu }$ be the four-velocity of a typical galaxy, satisfying the
condition $u^{\mu }u_{\mu }=-1$, with tetrad components $u^{a}=u^{\mu
}e_{\mu }^{a}$. The relativistic Boltzmann equation in tetrad components is
\begin{align}
u^{a}e_{a}^{\mu}\frac{\partial f_B}{\partial x^{\mu}}+
\gamma_{bc}^{a}u^{b}u^{c}\frac{\partial f_B}{\partial u^{a}}=0,  \label{tetr}
\end{align}
where the distribution function $f_B=f_B(x^{\mu },u^{a})$ and $\gamma
_{bc}^{a}=e_{\mu ;\nu }^{a}e_{b}^{\mu}e_{c}^{\nu}$ are the Ricci rotation
coefficients~\cite{Li66,Ja70}. By assuming that the only coordinate
dependence of the distribution function is upon the radial coordinate $r$,
Eq.~(\ref{tetr}) becomes~\cite{Ja70}
\begin{multline}
u_{1}\frac{\partial f_B}{\partial r}- \left(\frac{1}{2}u_{0}^{2}\frac{%
\partial \nu}{\partial r}- \frac{u_{2}^{2}+u_{3}^{2}}{r}\right) \frac{%
\partial f_B}{\partial u_{1}} \\
-\frac{1}{r}u_{1}\left( u_{2}\frac{\partial f_B}{\partial u_{2}} +u_{3}\frac{%
\partial f_B}{\partial u_{3}}\right) \\
-\frac{1}{r}e^{\lambda /2}u_{3}\cot \theta \left( u_{2} \frac{\partial f_B}{%
\partial u_{3}}-u_{3} \frac{\partial f_B}{\partial u_{2}}\right) = 0.
\label{tetr1}
\end{multline}

\subsection{The generalized virial theorem in $f(R)$ modified theories of
gravity}

The spherically symmetric nature of the problem requires that the
coefficient of $\cot \theta $, in Eq.~(\ref{tetr1}), be zero, which implies
that the distribution function $f_B$ is only a function of $r$, $u_{1}$ and $%
u_{2}^{2}+u_{3}^{2}$. Now, multiplying Eq.~(\ref{tetr1}) by $m u_{r} du $,
then integrating over the velocity space, and by assuming that $f_B$
vanishes sufficiently rapidly as the velocities tend to $\pm \infty $, we
obtain
\begin{multline}
r\frac{\partial}{\partial r}\left[\rho\left\langle u_{1}^{2}\right\rangle%
\right]+ \frac{1}{2}\rho \left[ \left\langle u_{0}^{2}\right\rangle +
\left\langle u_{1}^{2}\right\rangle\right] r\frac{\partial \nu }{\partial r}
\\
-\rho \left[ \left\langle u_{2}^{2}\right\rangle +\left\langle
u_{3}^{2}\right\rangle -2\left\langle u_{1}^{2}\right\rangle \right] =0.
\label{tetr2}
\end{multline}

The following step consists in multiplying Eq.~(\ref{tetr2}) by $4\pi r^{2}$%
, and integrating over the cluster provides the following relationship~\cite
{Ja70}
\begin{multline}
\int_{0}^{R}4\pi \rho \left[ \left\langle u_{1}^{2}\right\rangle
+\left\langle u_{2}^{2}\right\rangle +\left\langle u_{3}^{2}\right\rangle%
\right] r^{2}dr \\
-\frac{1}{2}\int_{0}^{R}4\pi r^{3}\rho \left[ \left\langle
u_{0}^{2}\right\rangle +\left\langle u_{1}^{2}\right\rangle\right] \frac{%
\partial \nu }{\partial r}dr=0.  \label{kin}
\end{multline}

It is convenient to introduce some approximations that apply to test
particles in stable circular motion around galaxies, and to the galactic
clusters. First of all, we assume that $\nu $ and $\lambda $ are slowly
varying (i.e.~$\nu^{\prime}$ and $\lambda^{\prime}$ small), so that in Eq.~(%
\ref{ff1}) the quadratic terms can be neglected. Secondly, we assume that
the galaxies have velocities much smaller than the velocity of the light, so
that $\langle u_{1}^{2}\rangle \approx \langle u_{2}^{2}\rangle \approx
\langle u_{3}^{2}\rangle \ll \langle u_{0}^{2}\rangle \approx 1$. Thus,
Eqs.~(\ref{ff1}) and (\ref{kin}) become
\begin{align}  \label{fin1}
\frac{1}{2r^{2}}\frac{\partial }{\partial r}\left(r^{2} \frac{\partial \nu }{%
\partial r}\right) = 4\pi G\rho + 4\pi G\rho_{\phi},
\end{align}
and
\begin{align}
2K-\frac{1}{2}\int_{0}^{R}4\pi r^{3}\rho \frac{\partial \nu }{\partial r}%
dr=0,  \label{cond1}
\end{align}
respectively, where
\begin{align}
K=\int_{0}^{R}2\pi \rho \left[ \left\langle u_{1}^{2}\right\rangle
+\left\langle u_{2}^{2}\right\rangle +\left\langle u_{3}^{2}\right\rangle %
\right] r^{2}dr,
\end{align}
is the total kinetic energy of the galaxies. The total mass of the system is
given by $M=\int_{0}^{R}dM(r)=\int_{0}^{R} 4\pi \rho r^{2}dr$. Note that the
main contribution to $M$ is due to the baryonic mass of the intra-cluster
gas and of the stars, but other particles, such as massive neutrinos, may
also contribute significantly to $M$.

Now, multiplying Eq.~(\ref{fin1}) by $r^{2}$ and integrating from $0$ to $r$
we obtain
\begin{align}
GM(r)=\frac{1}{2}r^{2}\frac{\partial \nu }{\partial r}-GM_{\phi }\left(
r\right),  \label{fin2}
\end{align}
where we have defined
\begin{align}  \label{darkmass}
M_{\phi }\left( r\right) =4\pi \int_{0}^{r}\rho _{\phi}(r')r'^{2}
dr'.
\end{align}

We may call this useful quantity as the \textit{geometric mass} of the
cluster. By multiplying Eq.~(\ref{fin2}) with $dM(r)$, following an
integration from $0$ to $R$, and by defining
\begin{align}
\Omega =-\int_{0}^{R}\frac{GM(r)}{r}\,dM(r),
\end{align}
and
\begin{align}
\Omega _{\phi }=\int_{0}^{R}\frac{GM_{\phi }(r)}{r}\,dM(r),
\end{align}
we obtain the following relationship
\begin{align}
\Omega =\Omega _{\phi }-\frac{1}{2}\int_{0}^{R}4\pi r^{3}\rho \frac{\partial
\nu }{\partial r}\,dr,
\end{align}
where the quantity $\Omega $ is the usual gravitational potential energy of
the system.

Finally, with the use of Eq.~(\ref{cond1}), we obtain the generalization of
the virial theorem, in $f(R)$ modified theories of gravity, which takes the
form
\begin{align}
2K + \Omega - \Omega_{\phi } = 0.  \label{theor}
\end{align}

Note that the generalized virial theorem, given by Eq.~(\ref{theor}), can be
written in an alternative form if we introduce the radii $R_{V}$ and $%
R_{\phi }$ defined by
\begin{align}
R_{V}=M^{2}\Bigl/\int_{0}^{R}\frac{M(r)}{r}\,dM(r),\Bigr.
\end{align}
and
\begin{align}
R_{\phi }=M_{\phi }^{2}\Bigl/\int_{0}^{R}\frac{M_{\phi }(r)}{r}\,dM(r),\Bigr.
\label{RU3}
\end{align}
respectively. We denote $R_{\phi }$ as the \textit{geometric radius} of the
cluster of galaxies. Thus, the quantities $\Omega$ and $\Omega _{\phi}$ are
finally given by
\begin{align}
\Omega =-\frac{GM^{2}}{R_{V}},
\end{align}
and
\begin{align}
\Omega _{\phi}=\frac{GM_{\phi }^{2}}{R_{\phi }},
\end{align}
respectively.

The virial mass $M_{V}$ is defined as \cite{JaH}
\begin{align}
2K=\frac{GMM_{V}}{R_{V}}.
\end{align}

After substitution into the virial theorem, given by Eq.~(\ref{theor}), we
obtain
\begin{align}  \label{fin6}
\frac{M_{V}}{M}=1+\frac{M_{\phi }^{2}R_{V}}{M^{2}R_{\phi }}.
\end{align}
If $M_{V}/M>3$, a condition which is valid for most of the observed galactic
clusters, then Eq.~(\ref{fin6}) provides the virial mass in $f(R)$ gravity,
which can be approximated by
\begin{align}
M_{V}\approx \frac{M_{\phi}^2}{M}\frac{R_{V}}{R_{\phi }}.  \label{virial}
\end{align}

The virial theorem can also be derived in the framework of
Newtonian gravity, and it can be formulated as \cite{Bi87,No}
\begin{equation}
\frac{d^{2}I_{jk}}{dt^{2}}=4K_{jk}+2W_{jk}+\frac{2}{3}\Lambda
I_{jk}+\ldots, \label{virNew}
\end{equation}
where $I_{jk}$ is the moment of inertia tensor of the gravitating system, $%
K_{jk}$ is the kinetic energy tensor, $W_{jk}$ is the
gravitational potential energy tensor and $\Lambda $ is the
cosmological constant. If an external force with potential $\Phi
_{\rm ext}$ acts on the system, a new term of the form
\begin{equation}
V_{jk}=-\frac{1}{2}\int \rho \left( x_{k}\frac{\partial \Phi _{\rm ext}}{%
\partial x_{j}}+x_{j}\frac{\partial \Phi _{\rm ext}}{\partial x_{k}}\right) dV,
\end{equation}
must be added to the right-hand side of Eq.~(\ref{virNew}). In
gravitational equilibrium, when $d^{2}I_{jk}/dt^{2}=0$, the virial
theorem gives $2K-\left| W\right| =0$, where $K$ and $W$ are the
traces of $K_{jk}$ and $W_{jk}$, respectively, and we have assumed
$\Lambda =0$. For $\left| W\right| =\left( 3/5\right)
GM_{V}^{2}/R_{V}$, corresponding to a constant density matter
distribution, we obtain that $K=\left( 3/10\right)
GM_{V}^{2}/R_{V}$. However, in the ``geometric mass'' approach to
the dark matter problem, in which one has to make a clear
distinction between matter, in the usual sense, and geometric
effects, only the baryonic matter has kinetic energy, which
requires a modification of the definition of the virial mass.
Moreover, the geometric effects, specific to $f(R)$ gravity
models, cannot be systematically considered in a purely Newtonian
derivation.

In the observational context, the virial mass $M_V$ is determined from the
study of the velocity dispersion $\sigma _r^2$ of the stars and of the
galaxies in the clusters. Now, according to our interpretation, most of the
mass in a cluster with mass $M_{tot}$ should be in the form of the geometric
mass $M_{\phi }$, so that $M_{\phi }\approx M_{tot}$. A possibility of
detecting the presence of the geometric mass and of the astrophysical
effects of the $f(R)$ extensions of general relativity is through
gravitational lensing, which can provide direct evidence of the mass
distribution and of the gravitational effects even at distances extending
far beyond of the virial radius of the galaxy cluster.

\section{Astrophysical applications}

\label{sec:d}

It is important to emphasize that astrophysical observations together with
cosmological simulations have shown that the virialized part of the cluster
corresponds roughly to a fixed density contrast $\delta \sim 200$ as
compared to the critical density of the universe, $\rho _{c}\left( z\right) $%
, at the considered redshift, so that $\rho _V=3M_{V}/4\pi R_{V}^{3}=\delta
\rho _{c}\left( z\right)$, where $\rho_V$ is the virial density, $M_{V}$ and
$R_{V}$ are the virial mass and radius; $\rho_{c}$ is given by $%
\rho_{c}\left( z\right) =h^{2}(z)3H_{0}^{2}/8\pi G$, where $h(z)$ is the
Hubble parameter normalized to its local value, i.e., $h^{2}(z)=\Omega
_{m}\left( 1+z\right) ^{3}+\Omega _{\Lambda }$, where $\Omega _{m}$ is the
mass density parameter and $\Omega _{\Lambda }$ is the dark energy density
parameter~\cite{Ar05}.

Now, once the integrated mass as a function of the radius is determined for
galaxy clusters, a physically meaningful fiducial radius for the mass
measurement has to be defined. The radii commonly used are either $r_{200}$
or $r_{500}$. These radii lie within the radii of the mean gravitational
mass density of the matter $\left\langle \rho _{tot}\right\rangle =200\rho
_{c}$ or $500\rho _{c}$. A pragmatic approach to the virial mass is to use $%
r_{200}$ as the outer boundary~\cite{ReBo02}. The numerical values of the
radius $r_{200}$ are in the range $r_{200}=0.85$ Mpc (for the cluster NGC
4636) and $r_{200}=4.49$ Mpc (for the cluster A2163), so that a typical
value for $r_{200}$ is approximately $2$ Mpc. The masses corresponding to $%
r_{200}$ and $r_{500}$ are denoted by $M_{200}$ and $M_{500}$, respectively,
and it is usually assumed that $M_{V}=M_{200}$ and $R_{V}=r_{200}$~\cite
{ReBo02}.

\subsection{Geometric mass and geometric radius from galactic cluster
observations}

Most of the baryonic mass in the clusters of galaxies is in the form of the
intra-cluster gas. The gas mass density $\rho _{g}$ distribution can be
fitted with the observational data by using the following expression for the
radial baryonic mass (gas) distribution~\cite{ReBo02}
\begin{align}
\rho _{g}(r)=\rho _{0}\left( 1+\frac{r^{2}}{r_{c}^{2}}\right) ^{-3\beta /2},
\label{dens}
\end{align}
where $r_{c}$ is the core radius, and $\rho _{0}$ and $\beta $ are
(cluster-dependent) constants.

It is usually assumed that the observed X-ray emission from the hot, ionized
intra-cluster gas is in isothermal equilibrium. Therefore, one may assume
that the pressure $P_{g}$ of the gas satisfies the equation of state $%
P_{g}=(k_{B}T_{g}/\mu m_{p})\rho_{g}$, where $k_{B}$ is Boltzmann's
constant, $T_{g}$ is the gas temperature, $\mu \approx 0.61$ is the mean
atomic weight of the particles in the cluster gas, and $m_{p}$ is the proton
mass~\cite{ReBo02}. Thus, with the use of the Jeans equation~\cite{Bi87} we
obtain the total mass distribution as~\cite{ReBo02,HaCh07}
\begin{align}
M_{tot}(r)=-\frac{k_{B}T_{g}}{\mu m_{p}G}r^{2}\frac{d}{dr}\ln \rho _{g}.
\end{align}

Now, taking into account the density profile of the gas given by Eq.~(\ref
{dens}), we obtain for the total mass profile inside the cluster the
following relation~\cite{ReBo02}
\begin{align}
M_{tot}(r)=\frac{3k_{B}\beta T_{g}}{\mu m_{p}G}\frac{r^{3}}{r_{c}^{2}+r^{2}}.
\label{mp}
\end{align}

According to the modified $f(R)$ gravity model, the total mass of
the cluster consists of the sum of the baryonic mass (mainly the
intra-cluster gas), and the geometric mass, so that
$M_{tot}(r)=4\pi \int_{0}^{r}\left( \rho _{g}+\rho _{\phi }\right)
r^{2}dr$. Hence it follows that $M_{tot}(r)$ satisfies the
following mass continuity equation,
\begin{align}
\frac{dM_{tot}\left( r\right) }{dr}=4\pi r^{2}\rho _{g}\left( r\right) +4\pi
r^{2}\rho _{\phi }\left( r\right).  \label{massf}
\end{align}

Since the gas density and the total mass profile inside the cluster are
given by Eqs.~(\ref{dens}) and (\ref{mp}), respectively, we can immediately
obtain the expression of the geometric density term inside the cluster as
\begin{align}
4\pi \rho _{\phi }\left( r\right)=\frac{3k_{B}\beta T_{g}\left(
r^{2}+3r_{c}^{2}\right) }{\mu m_{p}\left( r_{c}^{2}+r^{2}\right) ^{2}}-\frac{%
4\pi G\rho _{0}}{\left( 1+r^{2}/r_{c}^{2}\right) ^{3\beta /2}}.
\end{align}
In the limit $r\gg r_{c}$ we obtain for $\rho _{\phi }$ the simple relation
\begin{align}
4\pi \rho _{\phi }\left( r\right)=\left[ \frac{3k_{B}\beta T_{g}}{\mu m_{p}}%
-4\pi G\rho _{0}r_{c}^{3\beta }r^{2-3\beta }\right] \frac{1}{r^{2}}.
\end{align}

The geometric mass can be obtained generally as
\begin{multline}
  GM_{\phi}(r) = 4\pi\int_{0}^{r} r^{2} \rho _{\phi}(r)\,dr = 
  \frac{3k_{B}\beta T_{g}}{\mu m_{p}} \frac{r}{1+r_{c}^{2}/r^{2}} \\
  -4\pi G \rho_{0}\int_{0}^{r} 
  \frac{r^{2}dr}{\left(1+r^{2}/r_{c}^{2}\right)^{3\beta/2}},
\end{multline}
and in the limit $r\gg r_{c}$, may be approximated as
\begin{align}
GM_{\phi }\left( r\right) \approx \left[ \frac{3k_{B}\beta T_{g}}{\mu m_{p}}-%
\frac{4\pi G\rho _{0}r_{c}^{3\beta }r^{2-3\beta }}{3\left( 1-\beta \right) }%
\right] r.  \label{GM}
\end{align}

One may assume that the contribution of the gas density and mass to the
geometric density and geometric mass, respectively, can be neglected. The
latter approximations are very well supported by astrophysical observations,
which show that the gas represents only a small fraction of the total mass
\cite{ReBo02,Ar05,Sch01,Ca97}. Therefore, we obtain
\begin{align}\label{rhophi}
4\pi G\rho _{\phi }(r)\approx \left( \frac{3k_{B}\beta T_{g}}{\mu m_{p}}%
\right) r^{-2},
\end{align}
and
\begin{align}  \label{Mphix}
GM_{\phi }\left( r\right) \approx \left( \frac{3k_{B}\beta T_{g}}{\mu m_{p}}%
\right) r,
\end{align}
respectively.

By using the obtained mass profile one can obtain the form of the
metric tensor component $e^{\nu }$ inside the cluster.   From Eq.
(\ref{fin1}) it follows that $\nu ^{\prime }$ must satisfy the
condition $r^{2}\nu ^{\prime }=2M_{tot}(r)\approx 2M_{\phi }\left(
r\right) $, which gives
\begin{equation}\label{metrcl1}
e^{\nu }\approx C_{\nu }r^{2s},
\end{equation}
where $C_{\nu }$ is an arbitrary integration constant, and we
defined
\begin{equation}
s=\frac{3k_{B}\beta T_{g}}{\mu Gm_{p}},
\end{equation}
for notational simplicity. As for the metric coefficient
$e^{-\lambda } $, we can assume that in the first approximation it
is given by its standard general relativistic form, $e^{-\lambda
}\approx 1-2GM_{tot}/r$, and is given by $e^{-\lambda }\approx
1-6k_{B}\beta T_{g}/\mu m_{p}=1-2Gs$. However, this approximation
may not be necessary correct inside the clusters of galaxies.

Hence, in modified $f(R)$ gravity models the metric inside a
galactic cluster can be directly obtained from astrophysical
observations.

One may also estimate an upper bound for the cutoff of the
geometric mass. The idea is to consider the point at which the
decaying density profile of the geometric density associated to
the galaxy cluster becomes smaller than the average energy density
of the Universe. Let the value of the coordinate radius at the
point where the two densities are equal to be $R_{\phi }^{(cr)}
$. Then at this point $\rho _{\phi }(R_{\phi}^{(cr)})=\rho_{univ}$, where $%
\rho _{univ}$ is the mean energy density of the universe. By assuming $%
\rho_{univ}=\rho _{c}=3H^{2}/8\pi G=4.6975\times10^{-30}h_{50}^{2}\;\mathrm{g%
}/\mathrm{cm}^{-3}$, where $H=50h_{50}\;\mathrm{km}/\mathrm{Mpc}/\mathrm{s}$~\cite{ReBo02}, we obtain
\begin{eqnarray}
R_{\phi }^{(cr)}&=&\left( \frac{3k_{B}\beta T_{g}}{\mu m_{p}G\rho _{c}}%
\right)^{1/2}  \notag \\
&=&91. 33\sqrt{\beta }\left( \frac{k_{B}T_{g}}{5\text{keV}} \right)
^{1/2}h_{50}^{-1}\mathrm{Mpc.}  \label{Rucr}
\end{eqnarray}

The total geometric mass corresponding to this value is
\begin{eqnarray}
M_{\phi }^{(cr)}&=&M_{\phi }\left( R_{\phi }^{(cr)}\right)  \notag \\
&=&4.83\times 10^{16}\beta ^{3/2}\left( \frac{k_{B}T_{g}}{5\text{keV}}%
\right) ^{3/2}h_{50}^{-1}M_{\odot }.
\end{eqnarray}
This value of the mass is consistent with the observations of the mass
distribution in the clusters of galaxies. However, according to $f(R)$
modified theories of gravity, we predict that the geometric mass and its
effects extends beyond the virial radius of the clusters, which is of the
order of only a few Mpc.

\subsection{Radial velocity dispersion in galactic clusters}

The virial mass can also be expressed in terms of the characteristic
velocity dispersion $\sigma _{1}$ as~\cite{Ca97}
\begin{align}
M_{V}=\frac{3}{G}\sigma _{1}^{2}R_{V}.
\end{align}
By assuming that the velocity distribution in the cluster is isotropic, we
have $\langle u^2\rangle=\langle u_1^2\rangle+\langle u_{2}^2\rangle+\langle
u_{3}^2\rangle=3\langle u_1^2\rangle=3\sigma _r^2$, where $\sigma _r^2$ is
the radial velocity dispersion. $\sigma _1$ and $\sigma _r $ are related by $%
3\sigma _1^2=\sigma _r^2$.

In order to derive the radial velocity dispersion relation for clusters of
galaxies in $f(R)$ gravity we start from Eq.~(\ref{tetr2}). Taking into
account that the velocity distribution is isotropic, we obtain
\begin{align}
\frac{d}{dr}\left( \rho \sigma _{r}^{2}\right) +\frac{1}{2}\rho \frac{d\nu }{%
dr}=0.  \label{veldis}
\end{align}

One may assume that inside the cluster the condition $e^{-\lambda}\approx 1$, 
to a first order of approximation, and in the limit of small velocities
the modified field equation, Eq.~(\ref{ff1}), may be integrated to yield
\begin{align}
r^{2}\nu ^{\prime }=2GM_{\phi }(r)+2GM(r)+2C,
\end{align}
where $C$ is an arbitrary constant of integration. Since from Eq.~(\ref{veldis}) we have $\nu ^{\prime }=-(2/\rho )d\left( \rho \sigma_{r}^{2}\right) /dr$, it follows that the radial velocity dispersion of the galactic clusters, in $f(R)$ gravity, satisfies the differential equation
\begin{align}
\frac{d}{dr}\left( \rho \sigma _{r}^{2}\right) =-\frac{GM_{\phi }(r)}{r^{2}}%
\rho (r)-\frac{GM(r)}{r^{2}}\rho (r)-\frac{C}{r^{2}}\rho (r),
\end{align}
and which in turn provides the following general solution
\begin{multline}
\sigma _{r}^{2}(r)=-\frac{1}{\rho }\int^{r} \biggl[ \frac{GM_{\phi
}(r')}{r'^{2}}\rho (r')+\frac{GM(r')}{%
r'^{2}}\rho (r') \\
+\frac{C}{r'^{2}}\rho (r')\biggr] dr'+\frac{C_{1}}{%
\rho },  \label{veldisf}
\end{multline}
where $C_{1}$ is an integration constant.

As an example of the application of Eq.~(\ref{veldisf}), consider the case
in which the density $\rho $ of the normal matter inside the cluster has a
power law distribution, so that
\begin{align}
\rho (r)=\rho _{0}r^{-\gamma },
\end{align}
with $\rho _{0}$ and $\gamma \neq 1,3$ positive constants. The corresponding
normal matter mass profile is $M(r)=4\pi \rho _{0}r^{3-\gamma }/\left(
3-\gamma \right) $. The geometric mass $GM_{\phi }\approx\left( 3k_{B}\beta
T_{g}/\mu m_{p}\right)r= q_{0}r$, where $q_0=\left( 3k_{B}\beta T_{g}/\mu
m_{p}\right)$, is linearly proportional to $r$, as has been shown in the
previous Section.

Therefore, for $\gamma \neq 1,3$, we obtain the following solution
\begin{align}
\sigma _{r}^{2}(r)=\frac{q_{0}}{\gamma }+\frac{2\pi G\rho _{0}}{\left(
\gamma -1\right) \left( 3-\gamma \right) }r^{2-\gamma }+\frac{C}{\gamma +1}%
\frac{1}{r}+\frac{C_{1}}{\rho _{0}}r^{\gamma }.
\end{align}
For the specific case of $\gamma =1$, we find
\begin{align}
\sigma _{r}^{2}(r)=q_{0}+\frac{C}{2r}-2\pi G\rho _{0}r\ln r+\frac{C_{1}}{%
\rho _{0}}r,
\end{align}
and for $\gamma =3$, we have
\begin{align}
\sigma _{r}^{2}(r)=\frac{q_{0}}{3}-\pi G\rho _{0}\left( \ln r+\frac{1}{4}
\right) \frac{1}{r^{4}}+\frac{C}{4}\frac{1}{r}+\frac{C_{1}}{\rho _{0}} r^{3}.
\end{align}

The observed data for the velocity dispersion in clusters of
galaxies are usually analyzed by assuming the simple form $\sigma
_{r}^{2}(r)=B/(r+b) $ for the radial velocity dispersion, with $B$
and $b$ constants. As for the density of the galaxies in the
clusters the relation $\rho \left( r\right) =A/r\left( r+a\right)
^{2}$, with $A$ and $a$ constants, is used. The data are then
fitted with these functions by using a non-linear fitting
procedure~\cite{Ca97}. For $r\ll a$, $\rho (r)\approx A/r$, while
for $r\gg a$, $\rho (r)$ behaves like $\rho (r)\approx A/r^{3}$.
Therefore the comparison of the observed velocity dispersion
profiles of the galaxy clusters and the velocity dispersion
profiles predicted by $f(R)$ modified theories of gravity may give
a powerful method to discriminate between the different
theoretical models.

\section{The Lagrangian of the $f(R)$ gravity model}\label{lagr}

The virial theorem in $f(R)$ gravity models, which leads to the
possibility of obtaining the metric tensor components inside the
cluster, also opens the possibility of directly obtaining the
Lagrangian $f(R)$ of the theory from astrophysical observations.
To obtain the Lagrangian we start from the field equations
(\ref{f1})-(\ref{f3}), which in the standard representation can be
written as
\begin{equation}
F^{\prime \prime }-\frac{1}{2}\left( \nu ^{\prime }+\lambda
^{\prime }\right) F^{\prime }-\frac{\nu ^{\prime }+\lambda
^{\prime }}{r}F=0, \label{f11}
\end{equation}
\begin{eqnarray}\label{f12}
&&\nu ^{\prime \prime }+\nu ^{\prime 2}-\frac{1}{2}\left( \nu
^{\prime
}+\lambda ^{\prime }\right) \left( \nu ^{\prime }+\frac{2}{r}\right) -\frac{2%
}{r^{2}}\left( 1-e^{\lambda }\right)  =\nonumber\\
&&-2\frac{F^{\prime \prime }}{F}+\left( \lambda ^{\prime }+\frac{2}{r}%
\right) \frac{F^{\prime }}{F},
\end{eqnarray}
\begin{equation}
f=Fe^{-\lambda }\left[ \nu ^{\prime \prime }-\frac{1}{2}\left( \nu
^{\prime }+\lambda ^{\prime }\right) \nu ^{\prime
}-\frac{2}{r}\lambda ^{\prime }+\left( \nu ^{\prime
}+\frac{4}{r}\right) \frac{F^{\prime }}{F}\right] , \label{f13}
\end{equation}
and
\begin{equation}
R=2\frac{f}{F}-3e^{-\lambda }\left\{ \frac{F^{\prime \prime
}}{F}+\left[
\frac{1}{2}\left( \nu ^{\prime }-\lambda ^{\prime }\right) +\frac{2}{r}%
\right] \frac{F^{\prime }}{F}\right\} ,  \label{f14}
\end{equation}
respectively, where we have neglected the contribution of the
baryonic matter (intra-cluster gas, stars, etc). By using for
$\exp(\nu )$ the expression given by Eq.~(\ref{metrcl1}), and by
assuming that inside the cluster $\exp(\lambda )=$constant,
Eq.~(\ref{f11}) gives a second order linear differential equation for $F(r)$,
\begin{equation}
r^{2}F^{\prime \prime }-srF^{\prime }-2sF=0,
\end{equation}
with the general solution
\begin{equation}
F(r)=C_{1}r^{q_{-}}+C_{2}r^{q_{+}},
\end{equation}
where $C_{1}$ and $C_{2}$ are arbitrary integration constants, and
$q_{\pm }=\left( 1+s\pm \sqrt{1+10s+s^{2}}\right)/2$. By
considering for $F$ only the monotonically decreasing solution we
obtain $F(r) =C_{1}r^{-q}$, where
$q=-q_{-}=\left[\sqrt{1+10s+s^{2}}-\left(1+s\right)\right]/2\geq0$.
With this form of the function $F(r)$, Eq.~(\ref{f12}) fixes
$\exp(\lambda )$ as
\begin{eqnarray}
e^{\lambda }&=&1+s\left( 2-s\right) -q\left( q+2\right)
   \nonumber\\
&=&1-8s-3s^{2}+2s\sqrt{1+10s+s^{2}}.
\end{eqnarray}

Eq.~(\ref{f13}) gives the radial coordinate dependence of the
Lagrangian $f$ as
\begin{equation}
f=2C_{1}e^{-\lambda }\left[ 4(1+s)-2(2+s)\sqrt{1+10s+s^{2}}\right]
r^{-q-2},
\end{equation}
while from Eq.~(\ref{f14}) we obtain for $R$ the expression
\begin{equation}
R=\frac{\left( 1+5s\right) \sqrt{1+10s+s^{2}}-1-34s-9s^{2}}{r^{2}}%
e^{-\lambda }.
\end{equation}

Finally, we obtain $f$ as a function of $R$ in the form
\begin{equation}\label{act}
f(R)=f_{0}R^{1+q/2},
\end{equation}
where
\begin{eqnarray}
&&f_{0}=2C_{1}\times \nonumber \\
&&\frac{4(1+s)-2(2+s)\sqrt{1+10s+s^{2}}}{\left[ \left( 1+5s\right)
\sqrt{1+10s+s^{2}}-1-34s-9s^{2}\right] ^{1+q/2}}e^{q\lambda
/2}.\nonumber\\
\end{eqnarray}

Therefore, once the main physical parameters of the gas in the
cluster, such as the gas temperature $T_g$ or the gas density
profile, described by the parameter $\beta $,  are known, the
action of the modified gravity model can be completely obtained
from observations.

The weak field limit of the $f(R)$ generalized gravity models has
been discussed recently, for star-like objects, in~\cite{Chiba}
and~\cite{Fa08}, respectively. By assuming that $f(R)$ is an
analytical function at the constant curvature $R_{0}$, that
$m_{\phi }r\ll 1$, where $m_{\phi }$ is the effective mass of the
scalar degree of freedom of the theory, and that the fluid is
pressureless, the post-Newtonian potentials $\Psi \left( r\right)
$ and $\Phi \left( r\right) $ are obtained for a metric of the
form $ds^{2}=-\left[ 1-2\Psi \left( r\right) \right] dt^{2}+\left[
1+2\Phi(r) \right] dr^{2}+r^{2}\left( d\theta ^{2}+\sin ^{2}\theta
d\varphi ^{2}\right)$. One may then find the behavior of $\Psi
\left( r\right)$ and $\Phi(r)$ outside the star. This analysis
leads to a value of $\gamma =1/2$ for the Post-Newtonian parameter
$\gamma $, which from Solar System observations is known to have a
value of $\gamma =1$. An analysis of a Lagrangian of the form
given by Eq.~(\ref{act}), denoted in~\cite{Chiba} as
$f(R)=(R/\alpha )^{1+\delta }$ has also been considered, and the
conclusion is that ``$\ldots$this analysis is incapable of determining
whether $f(R)=R^{1+\delta }$ gravity with $\delta \neq 1$
conflicts with Solar System tests''~\cite{Chiba}. Therefore, once
the main physical parameters of the gas in the cluster, like the
gas temperature $T_g$, or the gas density profile, described by
the parameter $\beta $, are known, $q$ can be calculated,  the
action of the modified gravity model can be completely obtained
from observations, and the viability/non-viability of the model
can be directly tested by using cluster of galaxy data, which may
offer an alternative to the Solar System tests.

Another problem facing the $f(R)$ gravity models is the problem of
the stability. On a time scale of $\tau \approx 10^{-26}$ s a
``fatal instability'' develops when $f''(R)<0$~\cite{Fa08}. For
the Lagrangian given by Eq.~(\ref{act}) we have
$f''(R)=f_0(q/2)(q/2+1)R^{q/2-1}> 0$. Therefore this type of
instability does not develop in the present model.

\section{Summary and Discussions}

\label{sec:e}

Cosmology has entered a `golden age' in which the rapid
development of increasingly high-precision data has turned it from
a more theoretically driven to an observationally based science.
Nevertheless, modern astrophysics and cosmology are plagued with
two severe difficulties, known as the dark energy and the dark
matter problems. One promising approach of theoretical research to
improve our understanding of these issues is modified gravity. In
particular, $f(R)$ modified theories of gravity challenge the need
for dark matter and for dark energy. Some of the models seem to
account for the late time acceleration of the universe, and viable
models seem to exist. They simultaneously account for the four
distinct cosmological phases, inflation, the radiation-dominated
and matter-dominated epochs, and the late-time accelerated
expansion, respectively. Moreover, these models also seem to be
consistent with cosmological structure formation observations.

In an astrophysical context, galactic dynamics of massive test
particles may also be understood without the need for dark matter
in the framework of $f(R) $ modified theories of gravity. It is
therefore of paramount interest to derive a generalized version of
the virial theorem, which plays an important role in many areas of
astrophysics, in the context of the modified theories of gravity.
The virial theorem is an extremely useful tool in stability
issues, and in the deduction of the mean density of astrophysical
objects such as galaxies, clusters and super clusters, and
consequently in the prediction of the total mass of the clusters
of galaxies.

In this work, we have analyzed the `dark matter' problem by
considering a generalized version of the virial theorem in the
framework of $f(R)$ modified theories of gravity. The virial
theorem was obtained by using a method based on the collisionless
Boltzmann equation. The additional geometric terms present in the
modified gravitational field equations provide an effective
contribution to the gravitational energy, which at the
galactic/extra-galactic level acts as an effective mass, playing
the role of the `dark matter'. The total virial mass of the
galactic clusters is mainly determined by the effective mass
associated to the new geometrical term, the geometrical mass. It
is important that the latter term may account for the well-known
virial theorem mass discrepancy in clusters of galaxies.

 In the present model there
is also a strict proportionality between the virial mass of the
cluster and its baryonic mass, a relation which can also be tested
observationally. Since galaxy clusters are ``dark'' matter
dominated
objects, the main contribution to their mass comes from the geometric mass $%
M_{\phi }$, so that with a very good approximation we have $M_{\phi }\approx
M_{V}\approx M_{tot} $. Therefore the virial theorem gives immediately the
following mass scaling relation
\begin{align}
M_{V}\approx M\frac{R_{\phi }}{R_{V}}.
\end{align}
This equation shows that the virial mass is proportional to the baryonic
(normal) mass of the cluster, and that the ratio of the total mass and of
baryonic mass is determined by a purely geometric quantity, the geometric
radius $R_{\phi }$. Hence the geometric radius of the cluster can be
determined from observations, once the virial and baryonic masses and the
virial radius, respectively, are known.

By assuming that $R_{\phi }\approx R_{\phi }^{(cr)}$, we obtain the
following relation between the virial and the baryonic mass of the cluster
\begin{equation}
M_{V}\approx 91.33\sqrt{\beta }\left( \frac{k_{B}T_{g}}{5\;\mathrm{keV}}%
\right) ^{1/2}h_{50}^{-1}\frac{M}{R_{V}(\mathrm{Mpc})}.
\end{equation}
For a cluster with gas temperature $T_{g}=5\times 10^{7}$ K,
$\beta =1/2$ and $R_{V}=2$ Mpc we obtain $M_{V}\approx 32M$, a
relation which is consistent with the astronomical observations
\cite{ReBo02}.

One of the important predictions of the $f(R)$ ``dark matter''
model is that the geometric mass and its effects extends beyond
the virial radii of the clusters. Generally, the virial mass $M_V$
is obtained from the observational study of the velocity
dispersions of the stars in the cluster. Due to the observational
uncertainties it cannot give a reliable estimation of the
numerical value of the total mass $M_g+M_{\phi }$ in the cluster.
However, a much more powerful method for the study of the total
mass distribution in clusters is the gravitational lensing of
light, which may provide direct evidence for the gravitational
effects at large distances from the cluster. The presence of
$f(R)$ modified gravity effects at large distances from the
cluster leads to significantly different lensing observational
signatures, as compared to the standard relativistic/dark matter
model case. The bending angle in the $f(R)$ models could be larger
than that predicted by the dark matter models. Therefore, the
observational study of the gravitational lensing could
discriminate between the different dynamical laws proposed to
model the motion of particles at the clusters of galaxy level, and
the standard dark matter models.

In conclusion, the generalized virial theorem in $f(R)$ gravity
might be an efficient tool in observationally testing the
viability of this class of generalized gravity models.

\acknowledgments We would like to thank the two anonymous
referees, whose comments and suggestions helped us to
significantly improve the manuscript. The work of TH was supported
by the RGC grant No.~7027/06P of the government of the Hong Kong
SAR. FSNL was funded by Funda\c{c}\~{a}o para a Ci\^{e}ncia e a
Tecnologia (FCT)--Portugal through the grant SFRH/BPD/26269/2006.


\begin{thebibliography}{99}
\bibitem{JaH} We would like to thank Prof. John C. Jackson for
pointing out to us this definition of the virial mass.
\end{thebibliography}
\end{document}